\documentclass[preprint,nofootinbib]{revtex4}%
\usepackage{amssymb}
\usepackage{amsfonts}
\usepackage{amsmath}
\usepackage[utf8]{inputenc}
\usepackage{fancyhdr}
\usepackage{amsmath}
\usepackage{graphicx}
\usepackage{hyperref}
\usepackage{subcaption}
\usepackage{url}
\usepackage[usenames]{color}%
\providecommand{\U}[1]{\protect\rule{.1in}{.1in}}
\providecommand{\U}[1]{\protect\rule{.1in}{.1in}}
\definecolor{blue}{rgb}{0,0,1}

\definecolor{red}{rgb}{1,0,0}

\begin{document}
\title{Slowly rotating black holes
modeled by Solv geometries}
\author{José Figueroa and Marcelo Oyarzo}

\affiliation{Departamento de F\'{\i}sica, Universidad de Concepci\'{o}n, Casilla,
160-C, Concepci\'{o}n, Chile.}
\begin{abstract}
We present a slowly rotating generalization of a black hole modeled by a Solv horizon geometry, in five dimensional General Relativity. A separable ansatz compatible with Einstein equations is proposed, which is integrated in terms of hypergeometric and exponential functions. The solution highly simplifies after requiring to maintain the same asymptotic behaviour as the non-rotating black hole and to have finite, non-vanishing charges. We compute the charges of the solution and also comment on its importance for studying the phase space of this sector of the theory.
\end{abstract}
\maketitle
\section{Introduction}
Black hole solutions in General Relativity with negative cosmological constant $\Lambda$ can have a horizon geometry given either by flat, spherical or hyperbolic $\left(  d-2\right)  $-dimensional Euclidean manifolds \cite{Mann}. The flat and hyperbolic case only exist when $\Lambda$ is negative. In such cases the horizons are non-compact, but by making appropriate identifications they can be compactified, obtaining the topology of a torus for the planar case, or of a Riemann surface with higher genus for the hyperbolic case. These black holes also admit a rotating extension with non-vanishing angular momenta \cite{Klemm:1997ea}.

In five dimension, the $3$-dimensional horizon geometries are either flat $E^{3}$, spherical $S^{3}$ or hyperbolic $H_{3}$ spaces. These are a subset of the eight Thurston geometries, which are the building blocks of $3$-dimensional Euclidean spaces, as stated by the Geometrization conjecture \cite{Thurston:1997}. The
remaining five geometries are denoted by $S^{2}\times E$, $H_{2}\times E$,
$Nil$, $Sol$ and $SL\left(  2,R\right)  $. Interestingly, General Relativity in five dimension with negative $\Lambda$ also admits black holes with horizon geometry given by some of the previously mentioned Thurston geometries \cite{cadeu woolgar:2001}. One of them is the Sol black hole, whose metric is
\begin{equation}
ds^{2}=-\left(  \frac{-2\Lambda}{9}r^{2}-\frac{2\mu}{r}\right)  dt^{2}  + \left(  \frac{-2\Lambda}{9}r^{2}-\frac{2\mu}{r}\right)^{-1}dr^{2}%
+\frac{3 r^{2}}{-\Lambda}\left(  e^{2z}dx^{2}+e^{-2z}%
dy^{2}\right)  +\frac{3}{-\Lambda}dz^{2}\ , \label{solbh}%
\end{equation}
where $\mu$ is an integration constant related to the mass. 
When $\mu = 0$ one finds a kind of vacuum for this sector of the theory. Spacetimes approaching to that vacuum at infinity belong to a different phase space than those which are asymptotically AdS. Then, arises the natural question of whether the phase space of the Sol black holes can be enlarged, including, for example, rotating Sol black holes. In this work we constructed a new deformation of the black hole (\ref{solbh}) at the linear level, representing a slowly rotating solution, and we compute its charges using the covariant phase space formalism. We also find a linear hair of gravitational origin. 

\section{Sol Black hole and Vacuum}
Einstein equations in five dimensions admit the metric (\ref{solbh}) as a solution. As stated before, this solution represents a black hole, with event horizon located at $r=r_{+}=\left(  -9\mu/\Lambda \right)  ^{1/3}$. The geometry of the horizon
is modeled by the Sol geometry, namely, an homogeneous Euclidean space, with a
solvable isometry algebra. This spacetime can also be interpreted as an
inhomogeneous black string, since at each $z=cte$ slice the geometry of the
horizon gets modified.

Asymptotically the black hole goes as the geometry with $\mu=0$, namely
\begin{equation}
ds^{2}=-\left(  \frac{-2\Lambda}{9}r^{2}\right)  dt^{2}  + \left(  \frac{-2\Lambda}{9}r^{2}\right)^{-1}dr^{2}%
+\frac{3 r^{2}}{-\Lambda}\left(  e^{2z}dx^{2}+e^{-2z}%
dy^{2}\right)  +\frac{3}{-\Lambda}dz^{2}\ , \label{vaccum}%
\end{equation}
which is an Einstein manifold that is neither of constant curvature, nor conformally flat. This vacuum has a five dimensional isometry algebra generated by
\begin{align}
\xi_{\left(  1\right)  } =& -x\partial_{x}+y\partial_{y}+\partial
_{z} \nonumber ,\\
\xi_{\left(  2\right)  } =&-t\partial_{t}+r\partial_{r}-x\partial
_{x}-y\partial_{y} , \nonumber\\
\xi_{\left(  3\right)  }=&\partial_{t}, \quad \xi_{\left(  4\right)
}=\partial_{y},\quad \xi_{\left(  5\right)  }   =\partial_{x} \nonumber
\end{align}
with the following commutation relations
\begin{align*}
\left[  \xi_{1},\xi_{4}\right]   &  =-\xi_{4}\ ,\ \left[  \xi_{1},\xi
_{5}\right]  =\xi_{5}\\
\left[  \xi_{2},\xi_{3}\right]   &  =\frac{2}{3}\xi_{3}\ ,\ \left[  \xi
_{2},\xi_{4}\right]  =\frac{2}{3}\xi_{4}\ ,\ \left[  \xi_{2},\xi_{5}\right]
=\frac{2}{3}\xi_{5}.
\end{align*}
The only isometry broken by the mass of the black hole is the one generated by $\xi_{\left(
2\right)  }$. Note also that this algebra is not semisimple, it has an Abelian ideal, so it may acquire a non-trivial central extension. Now, in order to understand better the family of metrics that approach the background (\ref{vaccum}), we turn to the construction of the slowly rotating black hole.

\section{Slowly rotating Sol Black hole}
For planar Schwarzschild-AdS, the rotating solution can be obtained by a boost, since it has $\partial_{x}$ and $\partial_{y}$ as Killing vectors. Now we have the same symmetries, but even though the horizon of the Sol black hole is homogeneous, it has a non-vanishing
curvature, and the slowly \textquotedblleft rotating" solution is not locally
related to the static one.

Therefore, we propose the following ansatz for the slowly rotating metric
\begin{equation}
    ds^{2}=-f\left(  r\right)  dt^{2}+\frac{dr^{2}}{f\left(  r\right)  }  + d\tilde{s}^{2}   +a
F_{-}\left(  r\right)  G_{-}\left(  z\right)  dtdx+ b F_{+}\left(
r\right)  G_{+}\left(  z\right)  dtdy\ .
\label{ansatz}
\end{equation}
with
\begin{equation*}
   d\tilde{s}^{2} = r^{2} \left(  e^{2z} dx^{2}+e^{-2z} dy^{2}\right) + dz^{2}\ 
\end{equation*}
and where we have set $\Lambda=-3$. Interpreting this as a perturbation of the static black hole one can obtain,
in the presence of a Maxwell field, the response properties of the dual system \cite{Arias:2017yqj}.

Interestingly, Einstein equations are compatible with the separation we have assumed, and of course at leading order the metric function $f(r)$ is the same as in the static Black hole.

At linear order in the rotation parameters $a$ and $b$, we obtain an equation involving both pair of functions $F_{\pm}(r)$ and $G_{\pm}(z)$, that can be separated finding
\begin{equation}
\frac{d^{2}F_{\pm}\left(  r\right)  }{dr^{2}}-\frac{\left(  6f\left(
r\right)  -\left(  c_{\pm}^{2}-1\right)  r^{2}\right)  }{3f\left(  r\right)
r^{2}}F_{\pm}\left(  r\right)  =0\  
\label{eqFs}%
\end{equation}
and
\begin{equation}
\frac{d^{2}G_{\pm}\left(  z\right)  }{dz^{2}}\pm2\frac{dG_{\pm}\left(
z\right)  }{dz}-\left(  c_{\pm}^{2}-1\right)  G_{\pm}\left(  z\right)  =0\ ,
\end{equation}
where $c_{+}$ and $c_{-}$ are separation constants.
The slowly rotating Kerr metric can be obtained in the same way, but leads to slightly different equations. In this case, the equation for $G_{\pm}(z)$ can be solved as 
\begin{equation}
G_{\pm}\left(  z\right)  =C_{1,\pm}e^{\mp\left(  1+c_{\pm}\right)  z}%
+C_{2,\pm}e^{\mp\left(  1-c_{\pm}\right)  z}\,
\end{equation}
with $C_{1,\pm}$ and $C_{2,\pm}$ integration constants. The equations for $F_{\pm}(r)$ can be integrated in terms of hypergeometric functions, leading to the following asymptotic behaviors
\begin{equation}
F_{\pm}=D_{1,\pm}r^{\frac{1}{2}+\frac{1}{2}\sqrt{15-6c_{\pm}^{2}}}\left(
1+\mathcal{O}\left(  r^{-1}\right)  \right)  +D_{2,\pm}r^{\frac{1}{2}-\frac
{1}{2}\sqrt{15-6c_{\pm}^{2}}}\left(  1+\mathcal{O}\left(  r^{-1}\right)
\right)  \ .
\end{equation}
The branches with $D_{1,\pm}$ have to be discarded because they modify the
asymptotic behavior, and can even move us out from the perturbative approach, while the branches with $D_{2,\pm}$ go to zero at infinity
provided
\begin{equation}
-\frac{\sqrt{21}}{3}<c_{\pm}<\frac{\sqrt{21}}{3}\ . \label{allowed}%
\end{equation}

Then, the solution that decays at infinity written for arbitrary values of the
radial coordinate reads
\begin{equation}
F_{\pm}\left(  r\right)  =D_{2,\pm}r^{-\frac{3}{2}-\frac{1}{2}\sqrt
{15-6c_{\pm}^{2}}}f\left(  r\right)  \ F\left(  s_{\pm},s_{\pm}+1,2s_{\pm
},\frac{9\mu}{2r^{3}}\right)  \ ,
\end{equation}
where $F$ stands for a hypergeometric function $_2F_1$ and $s_{\pm}:=\frac{1}{2}%
+\frac{1}{6}\sqrt{5-6c_{\pm}^{2}}$.

In the next section we will see how analyzing the charges we can impose further constraints on the slowly rotating solution, leading to a considerable simplification of the metric.

\section{Charges and further constraints}
In order to obtain the charges we used the Covariant Phase Space Formalism. The expression for the variation of the charges is known \cite{Barnich:2001jy} and it is given by
\begin{equation}
\delta Q=\oint_{S} \mathbf{k}_{\xi}[\delta \Phi ; \Phi],
\end{equation}
where
\begin{equation}
\mathbf{k}_{\xi}^{\mu \nu}=\frac{\sqrt{-g}}{8 \pi G}\left(\xi^{\mu} \nabla_{\sigma} h^{\nu \sigma}-\xi^{\mu} \nabla^{v} h+\xi_{\sigma} \nabla^{v} h^{\mu \sigma}+\frac{1}{2} h \nabla^{\nu} \xi^{\mu}-\frac{1}{2} h^{\rho v} \nabla_{\rho} \xi^{\mu}+\frac{1}{2} h_{\sigma}^{\nu} \nabla^{\mu} \xi^{\sigma}\right).
\end{equation}

For our ansatz (\ref{ansatz}), we obtain the following charges
\begin{align}
\delta Q\left(  \partial_{t}\right)   &  =\frac{3\sqrt{3}}{4\pi
}\delta\mu\int dxdydz\ ,\\
\delta Q\left(  \partial_{x}\right)   &  =\lim_{r\rightarrow
\infty}\left(  -\frac{3\sqrt{3}}{32\pi}\delta a\int\left(
r\frac{\partial H_{-}\left(  r,z\right)  }{\partial r}-2H_{-}\left(
r,z\right)  \right)  rdxdydz\right)\ , \\
\delta Q\left(  \partial_{y}\right)   &  =\lim_{r\rightarrow
\infty}\left(  -\frac{3\sqrt{3}}{32\pi}\delta b\int\left(
r\frac{\partial H_{+}\left(  r,z\right)  }{\partial r}-2H_{+}\left(
r,z\right)  \right)  rdxdydz\right)\ .
\end{align}
Where $H_{\pm}(r,z)=F_{\pm}(r)G_{\pm}(z)$. Analyzing the integrand, we found that $c_{\pm}^{2}  \leq1$ in order to avoid divergences in $\delta Q\left(  \partial
_{x}\right)  $ and $\delta Q\left(  \partial
_{y}\right)  $ as $r\rightarrow\infty$.

For $c_{\pm}^{2}<1$, both charges $\delta Q\left(  \partial_{x,y}\right)  $
will vanish, and the off-diagonal term must be interpreted as a linear,
regular hair on the black hole background. Remarkably, only for $c_{\pm}%
^{2}=1$, the charges $\delta Q\left(  \partial_{x,y}\right)  $ receive a
non-vanishing contribution from the surface integral as $r\rightarrow+\infty$. Even more, in this case, in order to avoid
divergences in the $z$ direction one can turn off the corresponding integration constant in the $G_{\pm}(z)$ solution, giving rise to the following
the spacetime metric
\begin{equation}
ds^{2}=-f\left(  r\right)  dt^{2}+\frac{dr^{2}}{f\left(  r\right)  }%
+r^{2}\left(  e^{2z}dx^{2}+e^{-2z}dy^{2}\right)  +dz^{2}+\frac
{a}{r}dtdx+\frac{b}{r}dtdy\ ,
\label{final}
\end{equation}
which, defining $V_{T}=\int dxdydz$, has the following charges
\begin{align}
 Q\left(  \partial_{t}\right)   &  = M=\frac{3\sqrt{3}}{4\pi
}\mu V_{T}\\
 Q\left(  \partial_{x}\right)   &  = P_{x}=\frac{9\sqrt{3}}{32\pi
} a V_{T}\\
 Q\left(  \partial_{y}\right)   &  = P_{y}=\frac{9\sqrt{3}}{32\pi
} b V_{T}.
\end{align}

\section{Concluding remarks}
We have constructed new, slowly rotating solutions of General Relativity in five dimensions, with a negative cosmological constant. These represent the stationary extension of the static black hole found in \cite{cadeu woolgar:2001} with a horizon modeled by the Sol Thurston geometry. By computing the charges we have also shown that these solution contain a real parameter that is bounded from above. Below the bound, the off-diagonal terms lead to a hair of a gravitational origin since the only non-vanishing global charge is the mass of the spacetime. When the bound is saturated the spacetime acquires two extra non-vanishing global charges, namely $Q(\partial_x)$ and $Q(\partial_y)$.

It has been recently shown in \cite{Gutowski:2004ez} that the static Sol solution can be embedded in $\mathcal{N}=2$ gauged supergravity in five dimensions, in the presence of vector multiplets. Nevertheless such construction does not lead to supersymmetric black holes \cite{Gutowski:2004ez}. It would be interesting to explore whether the presence of a non-trivial rotation allows to by pass such result.

It is also interesting to notice that in the presence of matter fields without a supersymmetric origin, one can still construct black holes with Sol horizons, with dyonic sources \cite{Bravo-Gaete:2017nkp}. The construction of the rotating versions of such geometries is an open problem.

\section*{Acknowledgment}
The authors thank the organizers of  XXII Simposio Chileno de Física for the opportunity to present our work. Special thanks to Julio Oliva for his contribution and guidance. J.F and M.O also want to thank the financial support of Agencia Nacional de Investigaci\'on y Desarrollo (ANID) through the Fellowships No. 22191705. and No. 22201618  , respectively. This work is also partialy soported by FONDECYT grant 1181047.

\end{document}